\documentclass[manuscript,natbib209]{aastex}

\usepackage{longtable}
\usepackage{rotating}
\usepackage{multirow}
\usepackage{lscape}
\usepackage{color}
\usepackage{graphicx}

\shorttitle{LAMOST DR2 and 3 quasars.}
\shortauthors{*}

\begin{document}


\title{The Large Sky Area Multi-Object Fibre Spectroscopic Telescope (LAMOST) Quasar Survey: Quasar Properties from Data Release Two and Three}


\author{X. Y. Dong\altaffilmark{1}, Xue-Bing Wu\altaffilmark{1,2}, Y. L. Ai\altaffilmark{3}, J. Y. Yang\altaffilmark{1,2}, Q. Yang\altaffilmark{1,2}, F. Wang\altaffilmark{1,2}, Y. X. Zhang\altaffilmark{4}, A. L. Luo\altaffilmark{4}, H. Xu\altaffilmark{1,2}, H. L. Yuan\altaffilmark{4}, J. N. Zhang\altaffilmark{4}, M. X. Wang\altaffilmark{4}, L. L. Wang\altaffilmark{4}, Y. B. Li\altaffilmark{4}, F. Zuo\altaffilmark{4}, W. Hou\altaffilmark{4}, Y. X. Guo\altaffilmark{4}, X. Kong\altaffilmark{4}, X. Y. Chen \altaffilmark{4}, Y. Wu\altaffilmark{4}, H. F. Yang\altaffilmark{5}, M. Yang\altaffilmark{6} }

\altaffiltext{1}{Department of Astronomy, School of Physics, Peking University, Beijing,100871, P.R. China}
\altaffiltext{2}{Kavli Institute for Astronomy and Astrophysics, Peking University, Beijing,100871, P.R. China}
\altaffiltext{3}{School of Physics and Astronomy, Sun Yat-Sen University, Guangzhou 510275, China}
\altaffiltext{4}{Key Laboratory of Optical Astronomy, National Astronomical Observatories, Chinese Academy of Sciences 100012, Beijing, China}
\altaffiltext{5}{School of Computer Science and Technology, Taiyuan University of Science and Technology, Taiyuan 030024, China}
\altaffiltext{6}{National Observatory of Athens, Vas. Pavlou \& I. Metaxa, Penteli, 15236, Greece}

\email{sunne.xy.dong@pku.edu.cn}

\begin{abstract}

This is the second installment for the Large Sky Area Multi-Object Fibre Spectroscopic Telescope (LAMOST) Quasar Survey, which includes quasars observed from September 2013 to June 2015. There are 9024 confirmed quasars in DR2 and 10911 in DR3.  After cross-match with the SDSS quasar catalogs and NED, 12126 quasars are discovered independently. Among them 2225 quasars were released by SDSS DR12 QSO catalogue in 2014 after we finalised the survey candidates. 1801 sources were identified by SDSS DR14 as QSOs. The remaining 8100 quasars are considered as newly founded, and among them 6887 quasars can be given reliable emission line measurements and the estimated black hole masses. Quasars found in LAMOST are mostly located at low-to-moderate redshifts, with a mean value of 1.5. The highest redshift observed in DR2 and DR3 is 5. We applied emission line measurements to H${\sc \alpha}$, H${\sc \beta}$, Mg{\sc ii} and C{\sc iv}. We deduced the monochromatic continuum luminosities using photometry data, and estimated the virial black hole masses for the newly discovered quasars. Results are compiled into a quasar catalog, which will be available online. 

\end{abstract}





\section{Introduction}

Quasar is one of the most exotic celestial objects in the Universe. It emits a broad range of electromagnetic waves from $\gamma$-ray to radio \citep{Antonucci1993}, and outshines its host galaxy by hundreds of times. Quasars can be used to trace the galaxy evolution \citep[e.g.][]{Schweitzer2006, Lutz2008, Bonfield2011, Dong2016} and probe the intergalactic medium \citep[e.g.,][]{Hennawi2007, Huo2013, Huo2015}. 

There are many dedicated works to find more quasars ever since the first quasar was discovered by \cite{Schmidt1963}. Quasar's unique spectral energy distribution, its high luminosity and variations are often used to select quasar survey candidates by separating them from normal galaxies and stars. By far, the most productive quasar survey is the Sloan Digital Sky Survey \citep[SDSS;][]{Schneider2010, Paris2012, Paris2014}. Another noticeable quasar survey is the Two-Degree Fields (2dF) Quasar Redshift Survey \citep{Boyle2000}. Both SDSS and 2dF use ultra-violet/optical photometric data to select their candidates. With the improvements of both quality and quantity of infrared surveys, optical/infrared photometric data provide another important tool to select quasar candidates \citep[e.g.,][]{Wu2010, Wu2012, Ai2016}. The extreme-deconvolution method with an optical-Infrared color cut is used by the SDSS-III's Baryon Oscillation Spectroscopic Survey project \citep[XDQSO;][]{Bovy2011, Myers2015}. Other data-mining and machine-learning algorithms are also adopted for quasar selection \citep[e.g.,] []{Ross2012, Peng2012}.

The Large Sky Area Multi-Object Fibre Spectroscopic Telescope (LAMOST) Quasar Survey is conducted under the LAMOST ExtraGAlactic Survey \citep[LEGAS;][]{Zhao2012}. Its quasar candidates are selected based on multi-color photometry and data-mining. The regular survey started in September 2012 and will be carried out through five to six years. This paper is the second installment in the series of LAMOST quasar survey, after the pilot observations \citep{Wu2010b,Wu2010c} and data release 1 \citep[Paper I]{Ai2016}. We will report LAMOST quasar survey data release two and three, which include quasars observed between September 2013 and June 2015. In section 2, we will briefly review the survey and candidate selections. Section 3 describes the visual inspections of the spectra. The emission line measurements and the black hole mass estimates are discussed in section 4 and 5. The parameters released with the quasar catalog are list in section 6. The last section is the summary. Throughout the paper, we adopt a $\Lambda$-dominated cosmology with $h=0.7$, $\Omega_m$=0.3, and $\Omega_{\Lambda}$ =0.7. 

\section{Survey outline}

The Large Sky Area Multi-Object Fibre Spectroscopic Telescope (LAMOST), also known as Guoshoujing Telescope, is a 4-meter reflecting Schmidt telescope located at Xinglong Observatory, China \citep{Cui2012, Zhao2012}. It is equipped with 4000 fibres that across a $5^{\rm o}$ field of view. LAMOST has two sets of spectrographs. The blue channel covers spectrum from 3700${\rm\AA}$ to 5900${\rm\AA}$, while the red channel covers spectrum from 5700${\rm\AA}$ to 9000${\rm\AA}$. The spectral resolution $R$ is about 1800 over the entire wavelength range. 

\subsection{Target selections} 

The potential quasars candidates are required to be point sources in SDSS image survey \citep{Ahn2012} with a Galactic extinction corrected $i$-band Point Spread Function (PSF) magnitude brighter than 20mag. To avoid saturation and contamination from neighbour fibres, the magnitude upper-limit is set to be $i=16$. The final candidates are selected from this candidate pool via two methods.

{\bf Optical-infrared colors.} The foremost purpose of LEGAS quasar survey is to discover more quasars. Yet quasars at redshift 2 to 3 cannot be effectively identified in optical colour-colour space, because their colours are very similar to stars or galaxies \citep{Schneider2010}. \cite{Wu2010} and \cite{ Wu2012} proposed to use optical-infrared color instead. In the LEGAS survey we adopted $Y-K>0.46(g-z)+0.82$ or $J-k>0.45(i-Y-0.366)+0.64$ \citep{Wu2010} for candidates with counterparts within the UKIDSS/Large Area Survey(LAS) DR8 \citep{Lawrence2007}, and $w1-w2>0.57$ or $z-w1>0.66(g-z)+2.01$ \citep{Wu2012} for those found matches in WISE All-Sky Data Release \citep{Wright2010}. These selections mostly select quasars within redshift 4. 

{\bf Data-ming algorithms.} Some of the quasar candidates are selected using one of the following data-mining algorithms: support vector machine(SVM) classifiers \citep[SVM;][]{Peng2012}, extreme deconvolution method \citep{Bovy2011} and kernel density estimator \citep[KDE;][]{Richards2009}.

There are also some supplemented candidates selected based on multi-wavelength (optical/X-ray/radio) matching, such as X-ray sources from ROSAT, XMM-Newtown and Chandra, and radio sources from FIRST and NVSS. We included a group of SDSS DR7, DR9 and DR10 quasars on the purpose of testing LAMOST spectral liability. SDSS DR12 quasars are not removed because they were not released at the time when the candidate list was finalized for LAMOST. They are giving proper flags in the DR2 and DR3 quasar catalog. Some of our candidates overlap with the M31/M33 field \citep{Huo2015}. They are also flagged in the catalog. 

\subsection{Pipeline for data reduction}

The details of LAMOST data reduction can be found in \cite{Luo2015}. The raw data is reduced by the LAMOST 2D pipeline, which includes dark and bias removal, flat field correction, spectral extraction, sky subtraction and wavelength calibration. The output spectra are combinations of the blue and red channel spectra. 

The LAMOST 1D pipeline provides spectral type classification and radial velocity (for stars) or redshift (for galaxies and quasars) using the cross-correlation method. The primary classifications given by LAMOST 1D pipeline are STAR, GALAXY, QSO, and UNKNOWN. In general QSO and UNKNOWN have fainter magnitudes, therefore, visual inspection is required to confirm their spectral type and assure the redshift measurement. In the next section, we will discuss the visual inspection and compare the results with the pipeline results.

\section{Visual inspections of the spectra}

\begin{figure}
\includegraphics[scale=0.8]{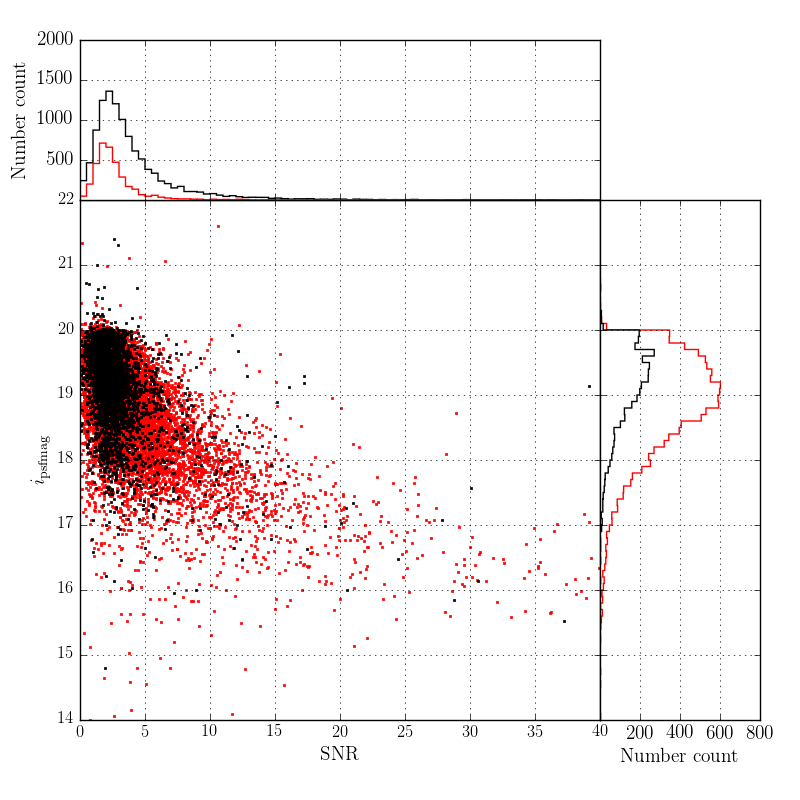} 
\caption{$i$-band PSF magnitude and the S/N of the spectrum. Red dots are pipeline QSOs. Black dots are pipeline UNKNOWNs but visually identified as quasars. The mean S/N of pipeline QSOs is  4.2, higher than the value of UNKNOWNs which is 2.8. The mean $i$-band magnitude of pipeline QSOs is 18.8 slight brighter than the UNKNOWNs' 19.1.  \label{Fig:snr-imag}}
\end{figure}

As mentioned in the previous section, 1D pipeline results are not trustworthy for QSO and UNKNOWN types. A Java program ASERA \citep{Yuan2013} is used to visually inspect each object with pipeline class QSO and quasar candidates with pipeline class UNKNOWN. The first step is to assure that the object is indeed a quasar. At the same time special objects such as BAL and disk emitter are labeled. In DR2, 9028 objects are identified as quasars, among them 2647 quasars are originally given UNKNOWN by the pipeline. Among 10911 confirmed DR3 quasars, 3428 quasars are previously UNKNOWN. There are still more than 12,000 quasar candidates with un-recognisable spectral type. They are returned to the candidates pool and will be re-observed in the future. The major reason to cause UNKNOWN type objects is the low S/N. It is either because the quasar candidates with magnitudes close to  $i=20$ are difficult for LAMOST to obtain good spectra, or due to the poor observational conditions at the time. Figure \ref{Fig:snr-imag} plots the distributions of the spectral S/N and the $i$-band PSF magnitude. It shows that the UNKONWN type spectra are fainter and have slightly lower S/N. If the blue and red channel spectra are improperly stacked, the combined spectrum might have artificial broad emission or absorption features, which also brings difficulties to pipeline classification.

\begin{figure}
\includegraphics[scale=0.8]{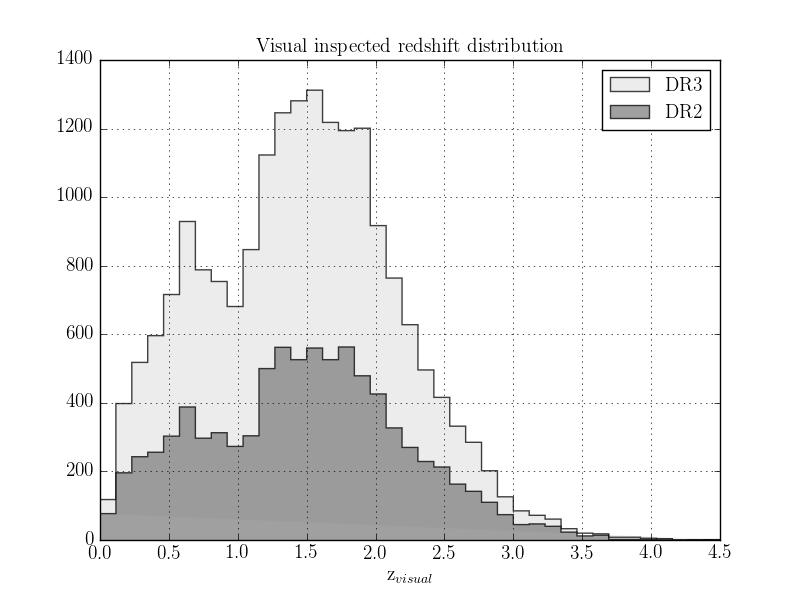} 
\caption{The visual inspected redshift distributions of DR2 (dark-grey) and DR3 (light-grey) quasars. The mean redshifts are 1.49 for DR2 and 1.43 for DR3, respectively. The quasar having highest redshift in DR2 is J132442.44+052438.8, with $z=5.01$. The quasar having highest redshift in DR3 is J112811.44+302255.9, with $z=4.90$. There are clear gaps in both DR2 and DR3 around $z\sim1$. It is mostly due to the artificial features at the spectral stacking region. 
\label{Fig:z-dist}}
\end{figure}

\begin{figure}
\includegraphics[scale=0.8]{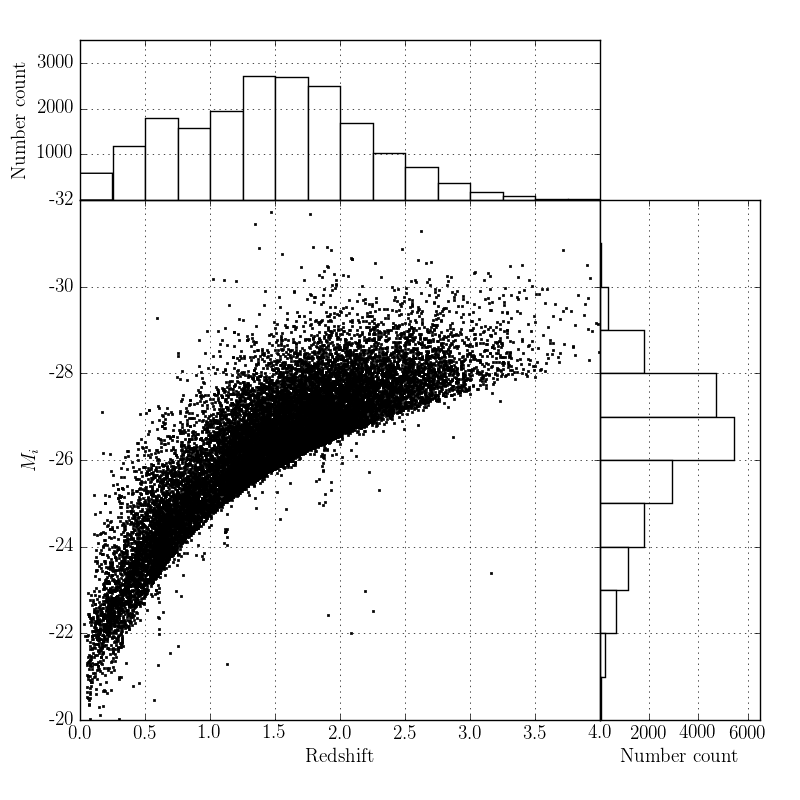} 
\caption{The redshift distribution on the luminosity space. The y-axis is the absolute $i$ band PSF magnitude with the Galactic extinction correction. The $K$-correction is normalised to $z=2$, following \cite{Richards2006}.
\label{Fig:z-lum}}
\end{figure}

The second step is to visually inspect redshift based on typical quasar emission lines, i.e., [O{\sc iii}], Mg{\sc ii}, C{\sc iii}, and C{\sc iv}. The poor quality of UNKNOWN type spectra also causes difficulties for pipeline to determine the redshift. For those with recognisable quasar emission lines, we visually estimate their redshifts using ASERA. But a flag will be given when there is only one emission line visualible. Sometimes, the visual inspected redshift mismatches the pipeline redshift. If there are at least two emission lines available, we manually correct the redshift, otherwise, we give a flag to indicate that the redshift for this object is arguable. Among visually confirmed quasars, 9484 have already been reported by SDSS. LAMOST pipeline gives redshifts of 6663 out of 9484, among them 95.7\% has redshift difference less than 0.2. After visual inspection, another 2440 quasar redshifts are recovered, among them more than 98\% have redshift differences less than 0.2 from SDSS. The redshifts of the remaining 381 SDSS-idenitified quasars cannot be measured using their LAMOST spectra of poor quality. Their redshift \textquotedblleft Z\_VI\textquotedblright\, in the catalog is 999 with Z\_FLAG = 9. Figure \ref{Fig:z-dist} is the visual inspected redshift distributions of DR2 and DR3. The mean redshifts of DR2 and DR3 is 1.49 and 1.44, respectively. The quasar having the highest redshift in DR2 and DR3 is J132442.44+052438.8, with a redshift of 5.01. When $z\sim1$, Mg{\sc ii} locates at the overlapped wavelength ranges (around 6000${\rm\AA}$) of the blue and red channels. If the spectral combinations of blue and red channels cause artificial features at the overlapping region, Mg{\sc ii} might become unrecognisable. It explains the missing $z\sim1$ objects in the distribution plots. Figure \ref{Fig:z-lum} plots DR2 and 3 quasars in the luminosity space, where $M_{i}$($z=2$) is $K$-corrected \citep{Richards2006} absolute $i$-band PSF magnitude normalised to $z=2$. 

\section{Spectral parameter measurements}

In this section we describe the procedures of typical quasar emission line measurement. Detailed discussions please refer to Paper I.

Before the measurement, the LAMOST spectrum is corrected with Galactic extinction using the extinction map of \cite{Schlegel1998} and the reddening curve of \cite{Fitzpatrick1999}. Then the spectrum is shifted back to the rest frame. 

Instead of fitting the continuum for the entire spectrum, we fit a local pseudo-continuum around each emission line \citep{Shen2008}. Two components are used for continuum fitting: a power law with its normalisation and slope as free parameters; and an Fe{\sc ii} mutliplet emissions with normalisation, velocity shift relative to the systemic redshift, and broadening velocity as free parameters. After subtraction of the local pseudo-continuum, the leftover of the spectrum is used for emission line fitting.

\subsection{H${\sc \alpha}$}

\begin{figure}
\includegraphics[scale=0.4]{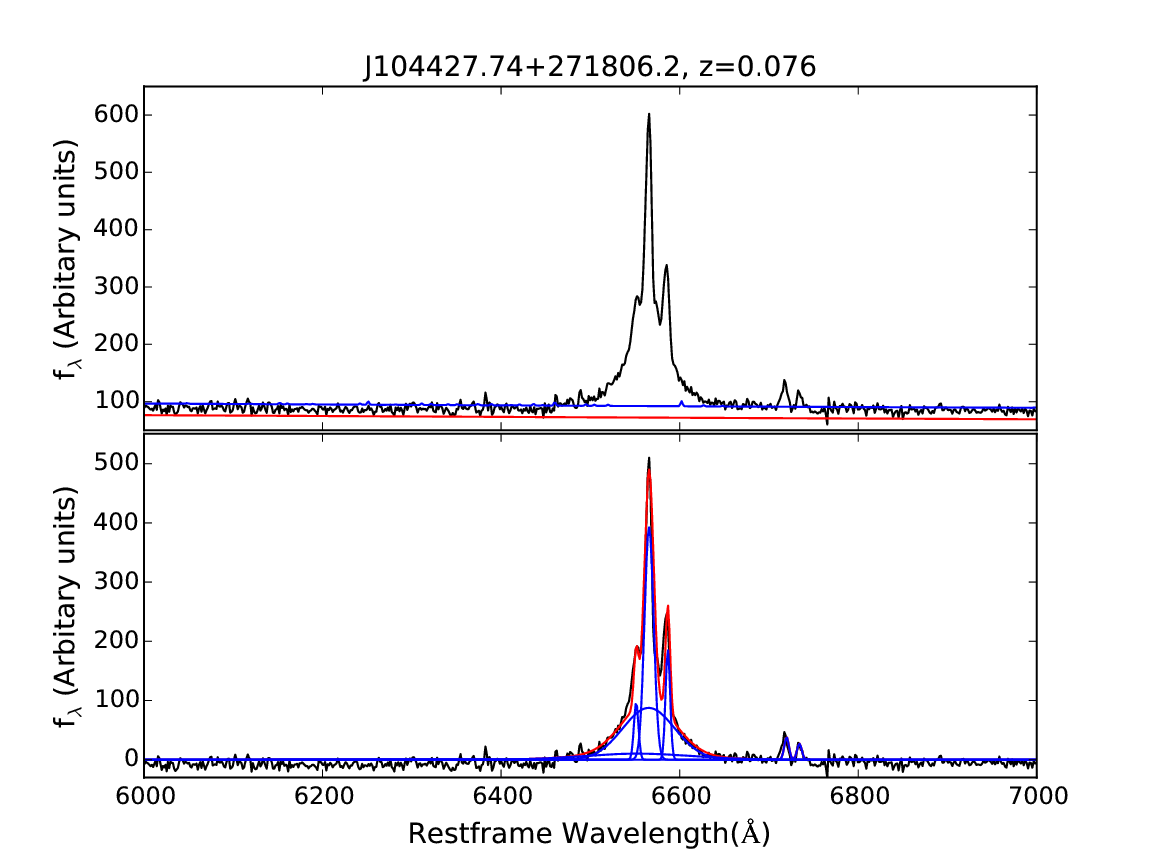} 
\caption{Example of H$\alpha$ measurement for J104427.74+271806.2, $z=0.076$. The top panel shows the pseudo-continuum fit. The red line is the power-law, the blue line is the power-law plus Fe{\sc ii} template. The bottom panel shows the H${\sc \alpha}$ emission line and [N{\sc ii}] fitting. The black line is the continuum subtracted spectrum. Each blue line represents one component. The red line is the combined emission line profile. 
\label{Fig:ha}}
\end{figure}

H${\sc \alpha}$ is only available when $z<=0.3$. The regions used for pseudo-continuum fitting are 6000-6250${\rm\AA}$ and 6800-7000${\rm\AA}$. The template of optical Fe{\sc ii} emission is from \cite{VeronCetty2004}. H${\sc \alpha}$ is modelled with one narrow and one broad component. The broad component is fitted with as many Gaussians as statistically justifiable (the typical number of Gaussian is four, see \cite{Dong2008} for detailed discussion). The narrow component is fitted with a single Gaussian with an upper limit of the full width half-maximum (FWHM) of 1200 km s$^{-1}$. The velocity offsets and line widths of [N{\sc ii}]$\lambda\lambda$6548, 6584 and [S{\sc ii}]$\lambda\lambda$6717,6731 are tied to H${\sc \alpha}$ narrow component. The relative flux ratio of the two [N{\sc ii}] components is fixed to 2.96. Figure \ref{Fig:ha} gives an example of H${\sc \alpha}$ fit.

\subsection{H${\sc \beta}$}

\begin{figure}
\includegraphics[scale=0.4]{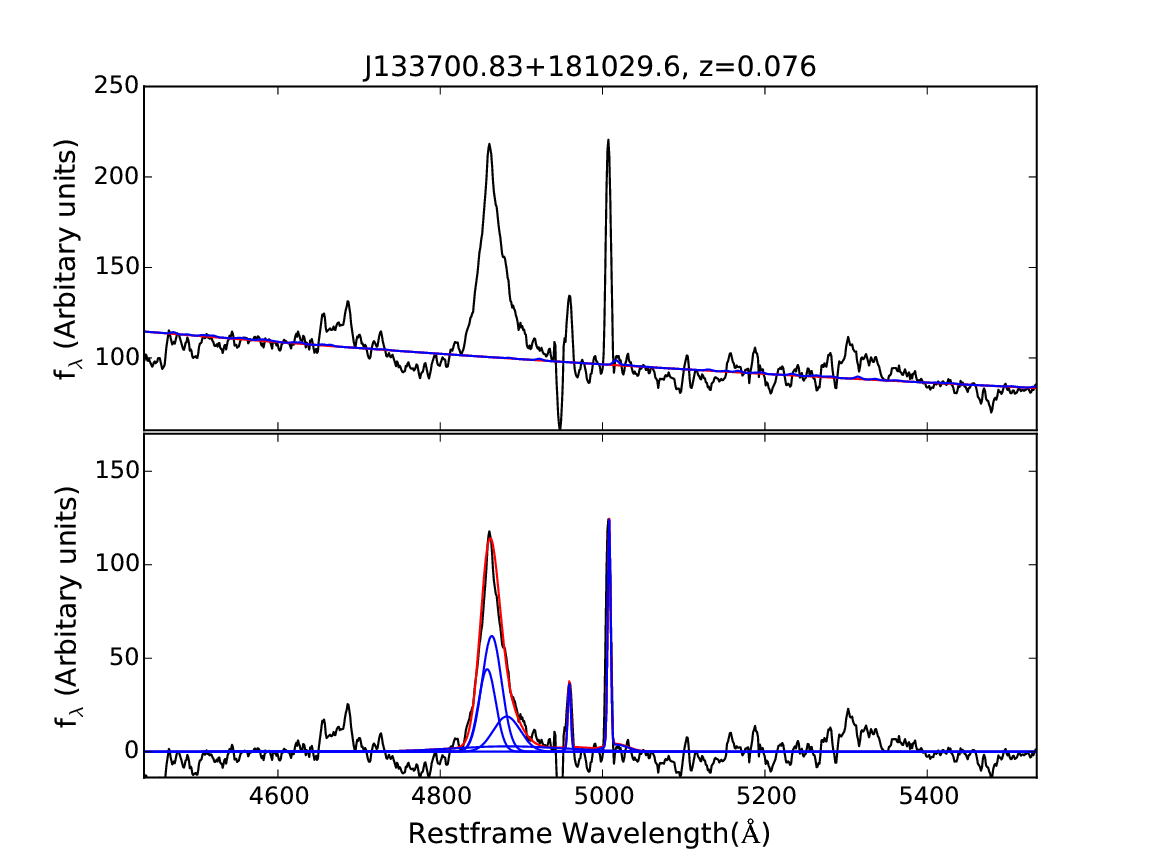} 
\caption{Example of H$\beta$ measurement for J133700.83+181029.6, $z=0.076$. The top panel shows the pseudo-continuum fit. The red line is the power-law, the blue line is the power-law plus Fe{\sc ii} template. The bottom panel shows the H${\sc \beta}$ emission line and [O{\sc iii}]. The black line is the continuum subtracted spectrum. Each blue line represents one component. The red line is the combined emission line profile.
\label{Fig:hb}}
\end{figure}

When $z<=0.75$, H${\sc \beta}$ is available. The regions for pseudo-continuum fitting are 4435-4700${\rm\AA}$ and 5100-5535${\rm\AA}$. The template of optical Fe{\sc ii} emission is from \cite{VeronCetty2004}. H${\sc \beta}$ is also modelled with a narrow and a broad component. The broad component is fitted with as many Gaussians as statistically justifiable \citep{Dong2008}. The narrow component is fitted with a single Gaussian with an upper limit of FWHM = 1200 km s$^{-1}$. For each line of the [O{\sc iii}]$\lambda\lambda$4959,5007 doublet, two Gaussians are used: one for the core and the other for a possible blue wing. The doublet are assumed to have same redshifts and profiles, with their flux ratio fixed to 3. The velocity offsets and line widths of the doublet core components are tied to H${\sc \beta}$ narrow component. Figure \ref{Fig:hb} gives an example of H${\sc \beta}$ fit.

\subsection{Mg{\sc ii}}

\begin{figure}
\includegraphics[scale=0.4]{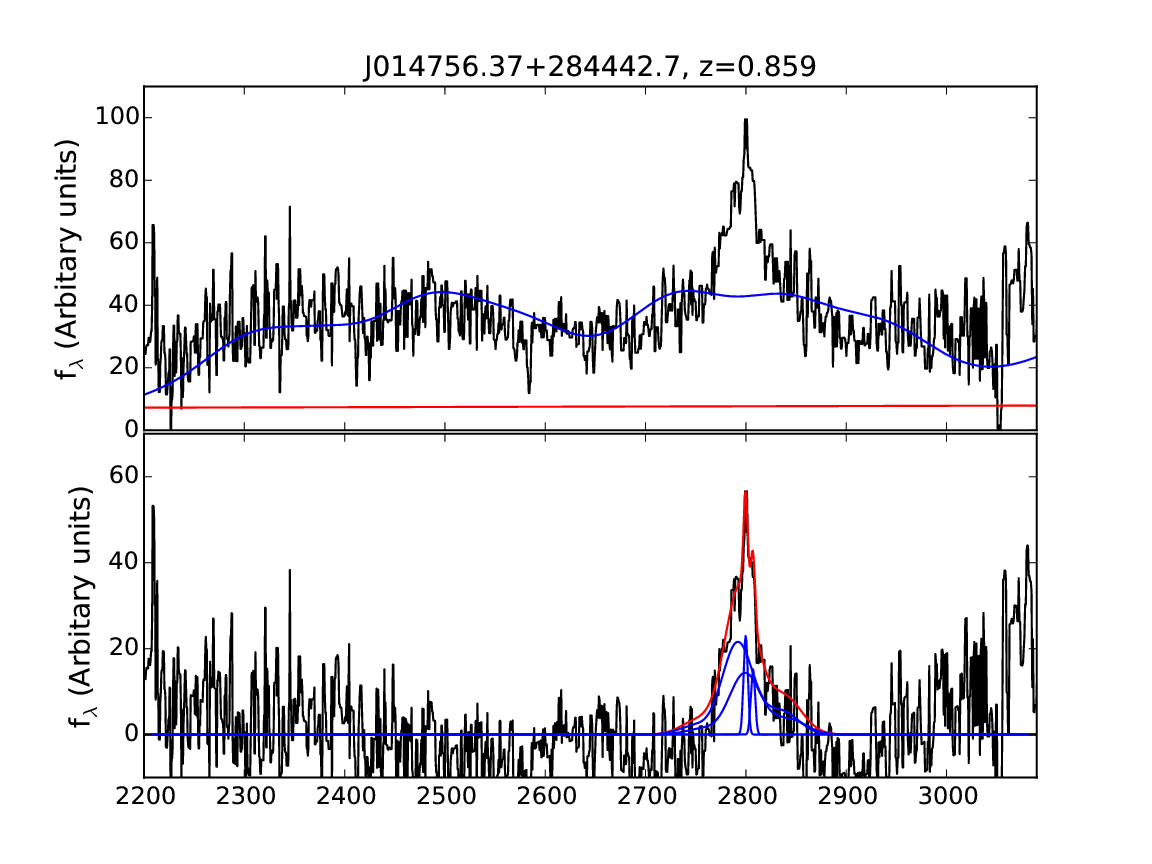} 
\caption{MgII measurement for J014756.37+284442.7, $z=0.859$. The top panel shows the pseudo-continuum fit. The red line is the power-law, the blue line is the power-law plus Fe{\sc ii} template. The bottom panel shows Mg{\sc ii} doublet fit. The black line is the continuum subtracted spectrum. The blue lines include one narrow and one broad components for each of the Mg{\sc ii} doublet. The red line is the combined Mg{\sc ii} profile. \label{Fig:mg}}
\end{figure}

Mg{\sc ii} is measured when quasar redshift is between 0.40 and 1.76. The pseudo-continuum is fitted using data from 2200-2700 ${\rm\AA}$ and 2900-3090${\rm\AA}$. An ultraviolet Fe{\sc ii} template from \cite{Tsuzuki2006} is included in the continuum fitting. One narrow component and one broad component are used to fit each of the Mg{\sc ii}$\lambda\lambda$2796,2803 doublet. The narrow component is a single Gaussian with a FWHM less than 900 km s$^{-1}$. The broad component is a truncated five-parameter Gauss-Hermite series. The broad component of each Mg{\sc ii} doublet line is assumed to have same profile with flux ratio set to be between 2:1 and 1:1 \citep{Laor1997}. The same assumption is applied to the narrow component of each Mg{\sc ii} doublet line. Figure \ref{Fig:mg} shows an example of Mg{\sc ii} fit. 

\subsection{C{\sc iv}}

\begin{figure}
\includegraphics[scale=0.4]{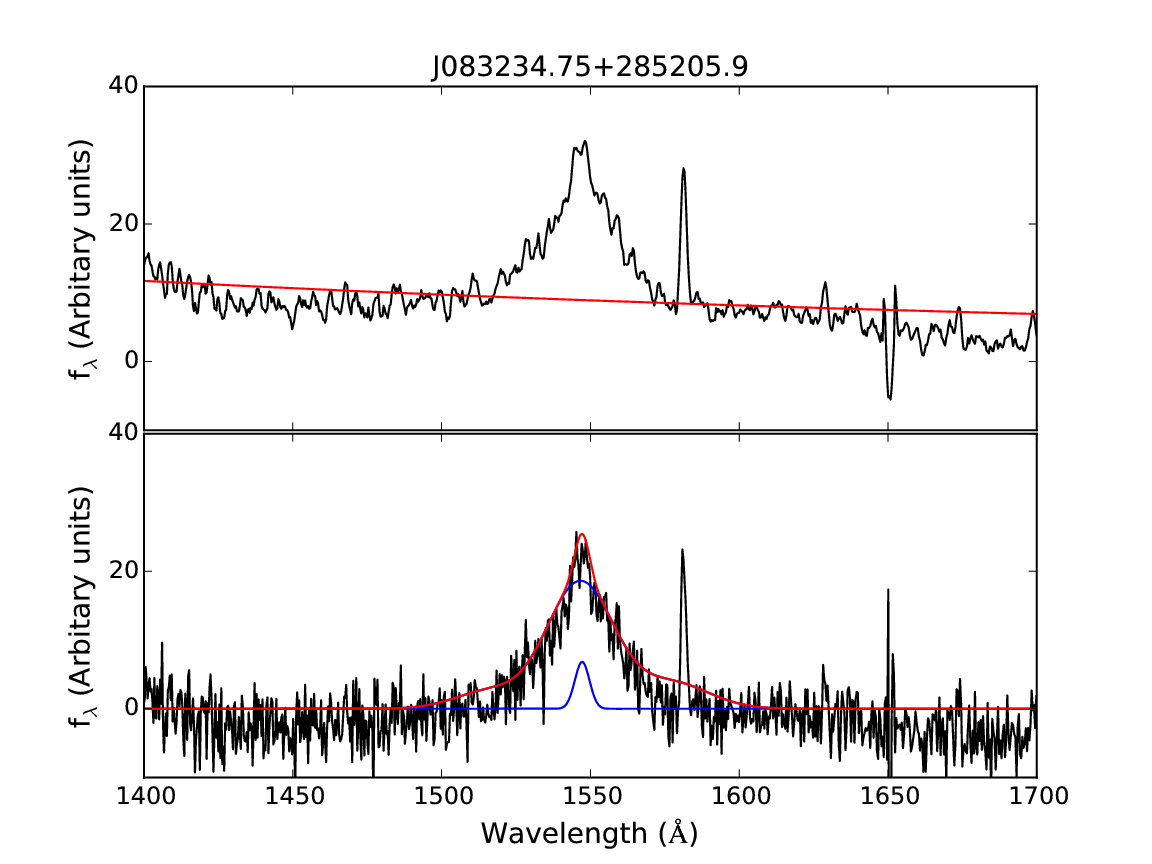} 
\caption{CIV measurement for J083234.75+285205.9, $z=2.34$. The top panel shows the pseudo-continuum fit. The red line is the power-law. The bottom panel shows the C{\sc iv} emission line measurement. The black line is the continuum subtracted spectrum. One of the blue line is a single Gaussian, the other blue line is a Gaussian-Hermite series. The red line is the combined emission line profile.  \label{Fig:civ}} 
\end{figure}

C{\sc iv} is measured when quasar redshift is larger than 1.58. The fitting regions for the pseudo-continuum are 1145-1465${\rm\AA}$ and 1700-1705${\rm\AA}$. One Gaussian and one Gaussian-Hermite are used to fit C{\sc iv} emission. There is no upper limit set for the FWHM of the Gaussian function, because the existence of a strong narrow component for C{\sc iv} is still debatable. An example of C{\sc iv} fitting is presented in Figure \ref{Fig:civ}.

\begin{figure}
\includegraphics[scale=0.4]{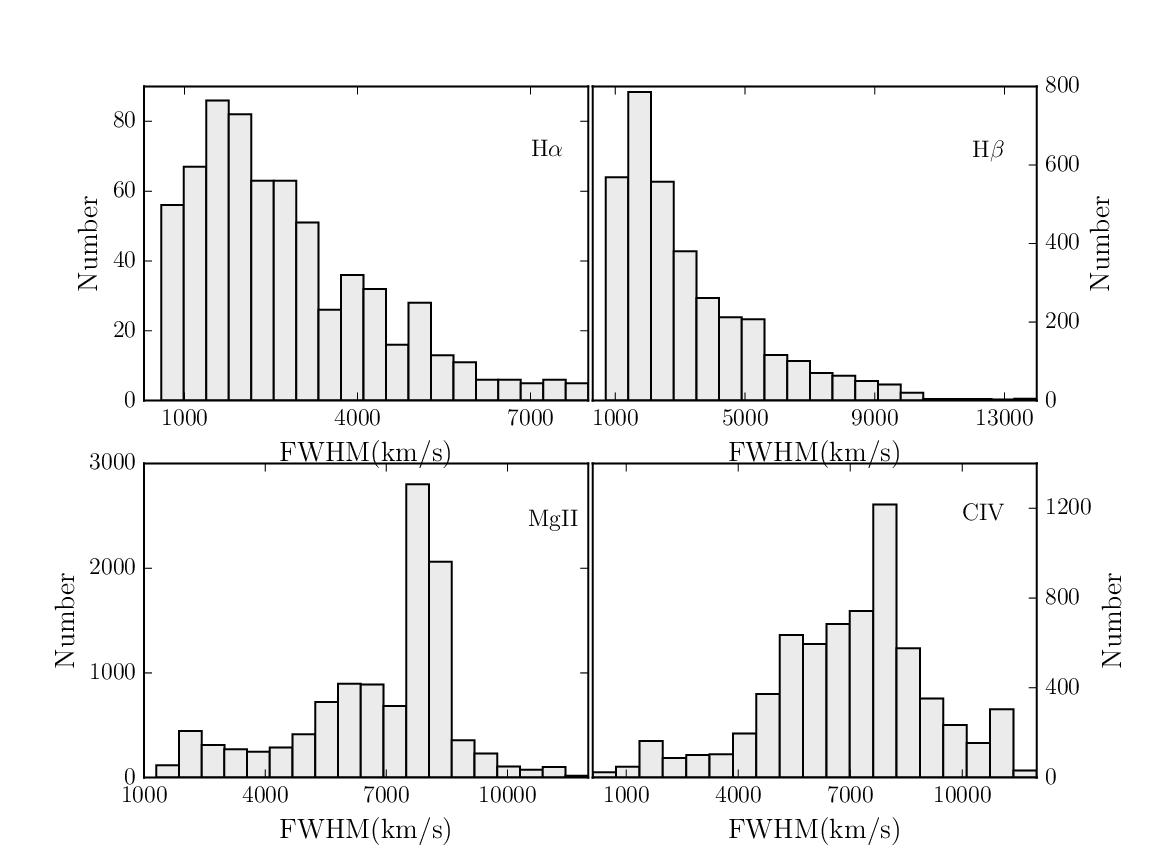} 
\caption{The distributions of FWHM of emission lines H${\sc \alpha}$, H${\sc \beta}$, Mg{\sc ii}, and C{\sc iv}.  \label{Fig:fwhm}} 
\end{figure}

The distributions of the FWHM of H${\sc \alpha}$, H${\sc \beta}$, Mg{\sc ii}, and C{\sc iv} are shown in Figure \ref{Fig:fwhm}. After spectra measurements, each fitting results are inspected visually. Flags are added based on the quality of the fit. 

\subsection{The reliability of the emission line fitting.}\label{em-fit}
\begin{figure}
\includegraphics[scale=0.8]{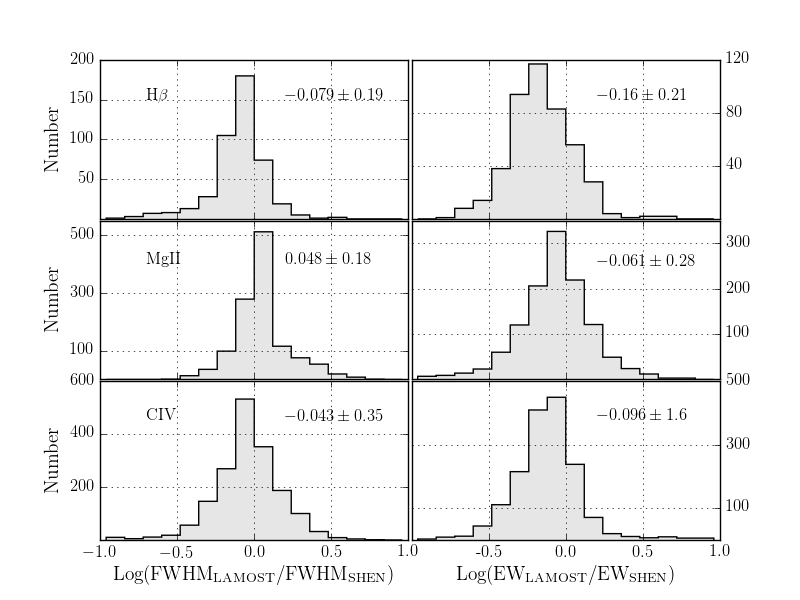} 
\caption{The comparisons of FWHM and EW between LAMOST DR2\&3 and \cite{Shen2011}. \label{Fig:comp-fwhm-ew}}
\end{figure}

As mentioned in Paper I, the core routines of the emission line fitting are adopted from \cite{Dong2008} and \cite{Wang2009}, which used the MPFIT package \citep{Markwardt2009} to perform $\chi^2$-minimization using the Levenberg-Marquardt technique. One may find the detailed discussion of the measurement uncertainties and justifications in \citet[][Section 2.5]{Dong2008} and \citet[][Section 2.2]{Wang2009}. \citet[][Section 3.3, thereafter Shen11]{Shen2011} also provides a detailed discussion regarding the different measurements of Mg{\sc ii}. To justify the fitting results of this set of data, we provide the simple comparisons between LAMOST DR2\&3 and that of Shen11.  Histograms on the left side of Figure \ref{Fig:comp-fwhm-ew} compare the FWHMs of H${\sc\beta}$, Mg{\sc ii}, and C{\sc iv}, while histograms on the right side compare their EWs. It appears that our measurements of FWHM and EW for H${\sc\beta}$ and C{\sc iv} are smaller than Shen11, with a mean difference of $-0.079\pm0.19$ dex and $-0.16\pm0.21$ dex for H${\sc\beta}$, and $=-0.043\pm0.35$ dex and $-0.10\pm1.6$. 
The FWHM of Mg{\sc ii} is larger than Shen11 and the EW is smaller, with a mean difference of $0.048\pm0.18$ dex and $-0.061+0.28$ dex respectively. As mentioned in previous sections, four Gaussians are used to fit the broad component of H${\sc\beta}$, and one Gaussian and one Gaussian-Hermite functions are used to fit the C{\sc iv} emission line in our work. In Shen11 a single or a maximum number of three Gaussians are used to fit the broad component of H${\sc\beta}$ and C{\sc iv}. For Mg{\sc ii}, we used doublets with one narrow and one broad components, while Shen11 used one or a maximum number of three Gaussians. Shen11 states that these two fitting procedures yield consistent results in FWHM of Mg{\sc ii} after applying both methods on the SDSS spectra. It is plausible that the small differences showed in Figure \ref{Fig:comp-fwhm-ew} are due to the differences in the fitting procedures and/or the different SNRs of LAMOST/SDSS spectra.

\section{Continuum luminosity and the virial black hole mass}\label{cont-fit}

\begin{figure}
\includegraphics[scale=0.4]{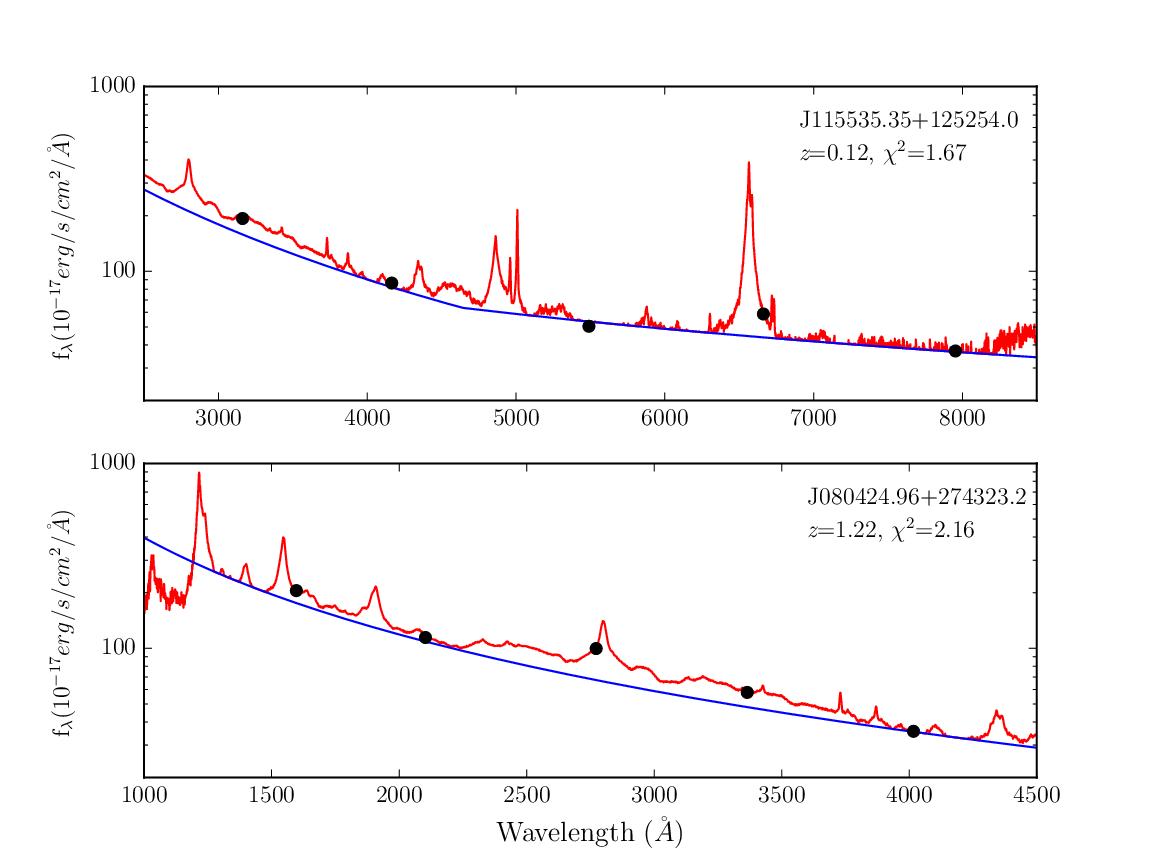} 
\caption{Two examples of continuum fitting. The x-axis is the rest-frame wavelength, the y-axis is the flux density. The black dots are 5-band SDSS photometry. The blue lines are the continuum represented by two power-laws. The red lines are composite spectra. The top panel is a quasar with $z=0.12$, its continuum has a clear break at 4661${\rm\AA}$. The bottom panel is a quasar with $z=1.22$, a single power-law is required for its continuum. \label{Fig:lumfit}} 
\end{figure}

\begin{figure}
\includegraphics[scale=0.4]{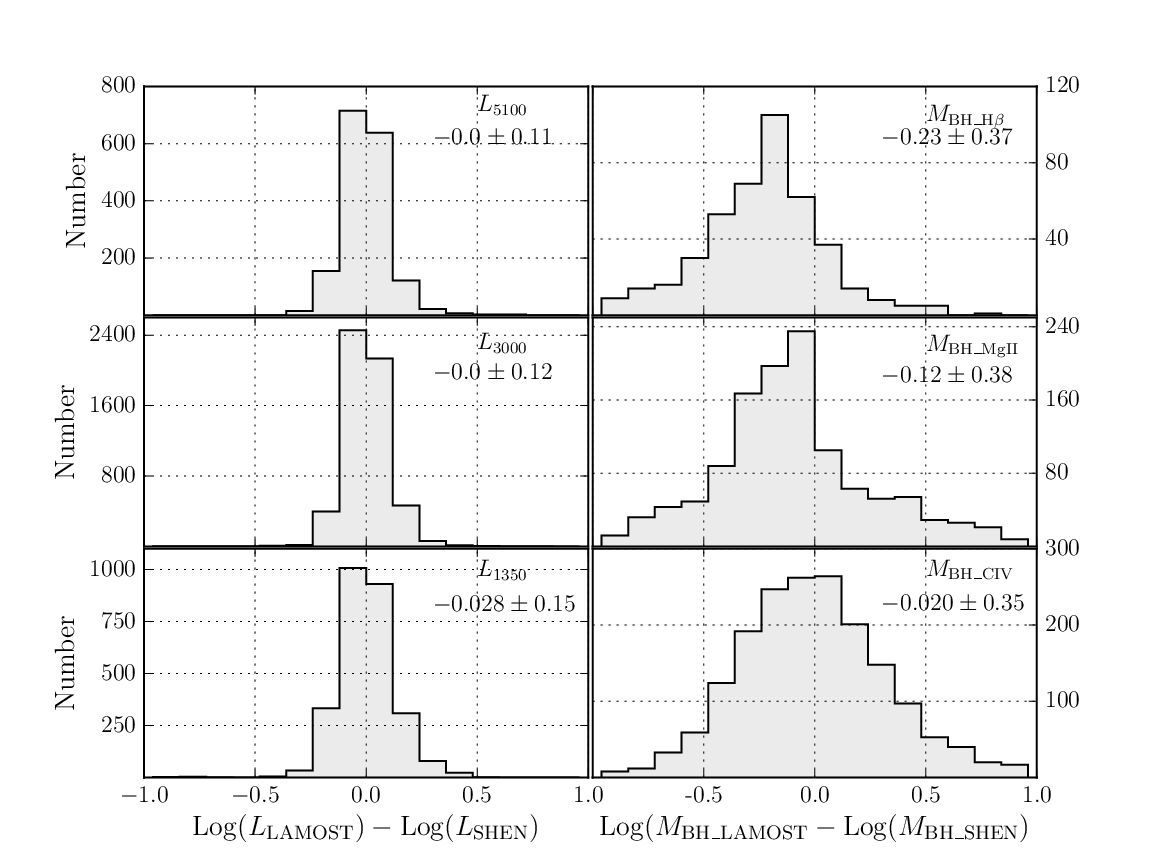} 
\caption{The comparisons of the monochromatic continuum luminosities and the estimated BH masses with those in \cite{Shen2011}. \label{Fig:comp-bol-BH}}
\end{figure}

The LAMOST survey is not equipped with a photometry telescope \citep{Song2012}. Its large field of view makes it difficult in some cases to find a suitable flux standard star for each spectrograph especially for those relatively faint extra-galactic objects. Therefore, the released LAMOST quasar spectra either are not flux calibrated or have poor flux calibrations. To estimate continuum luminosity we use the five-band SDSS photometry instead. The PSF magnitudes of each quasar are retrieved from SDSS photometry database. Before the photometry fit, these magnitudes are corrected with Galactic extinction and converted to flux density, $F_\lambda$, where $\lambda$ is the effective wavelength of each filter. Two components are used to fit the photometry, i.e., a continuum spectrum and an emission line template. As described in \cite{VandenBerk2001}(hereafter VB01), the continuum spectrum can be represented by two power-laws with a break at 4661${\rm\AA}$. The spectral indices given by VB01 are $\alpha_\lambda=-1.54$ at the blueward of 4661${\rm\AA}$ and $\alpha_\lambda=-0.42$ at the redward of 4661 ${\rm\AA}$. The emission line template is generated by removing the continuum spectrum from the composite median quasar spectrum. The normalisations of the continuum spectrum and emission line template are free parameters. The other two free parameters are the spectral indices of the two power-laws, with $\alpha_\lambda=-1.54$ and $\alpha_\lambda=-0.42$ as their initial values, respectively. The fitting uses Python PYTOOLS.NMPFIT \footnote{http://cars.uchicago.edu/software/python/index.html} to perform $\chi^2$-minimization with the Levenberg-Marquardt algorithm. Figure \ref{Fig:lumfit} presents two luminosity fitting examples. The results are well matched with the composite spectrum\citep{VandenBerk2001}, with an average spectral index of -1.52 blueward of 4661${\rm\AA}$ and -0.33 redward of 4661${\rm\AA}$. The monochromatic continuum luminosities at 1350${\rm\AA}$, 3000${\rm\AA}$, and 5000${\rm\AA}$ are calculated from the power-law continuum spectra. We compared the monochromatic continuum luminosities to \cite{Shen2011} and drew the histograms on the left side of Figure \ref{Fig:comp-bol-BH}. It is clear that our results agree very well with \cite{Shen2011}.

We calculate virial black hole mass using H${\sc \beta}$, Mg{\sc ii} and C{\sc iv} based on \cite{Shen2011} scheme as follows:

\begin{itemize}
\item $M_{\rm BH}$(H${\sc \beta}$), \begin{equation}{{\rm Log_{10}}(\frac{M_{\rm BH,vir}}{M_{\odot}})=0.910+0.50 {\rm Log_{10}}(\frac{L_{5100}}{\rm 10^{44} erg\,s^{-1}})+2 {\rm Log_{10}}(\frac{\rm FWHM(H\beta)}{\rm km\,s^{-1}})}\label{eq:hbeta}\end{equation} \citep{Vestergaard2006}
\item $M_{\rm BH}$(Mg{\sc ii}), \begin{equation}{\mathrm {Log_{10}}(\frac{M_{\rm BH,vir}}{M_{\odot}})=0.740+0.62 {\rm Log_{10}}(\frac{L_{3000}}{\rm 10^{44} erg\,s^{-1}})+2 {\rm Log_{10}}(\frac{\rm FWHM({Mg\,{\scriptstyle II}})}{\rm km\,s^{-1}})}\label{eq:mgii}\end{equation} \citep{Vestergaard2009}
\item $M_{\rm BH}$(C{\sc iv}), \begin{equation}{{\rm Log_{10}}(\frac{M_{\rm BH,vir}}{M_{\odot}})=0.660+0.53 {\rm Log_{10}}(\frac{L_{1350}}{\rm 10^{44} erg\,s^{-1}})+2 {\rm Log_{10}}(\frac{\rm FWHM({C\,{\scriptstyle IV}})}{\rm km\,s^{-1}})}\label{eq:civ}\end{equation} \citep{Shen2010}.
\end{itemize}

\begin{figure}
\includegraphics[scale=0.8]{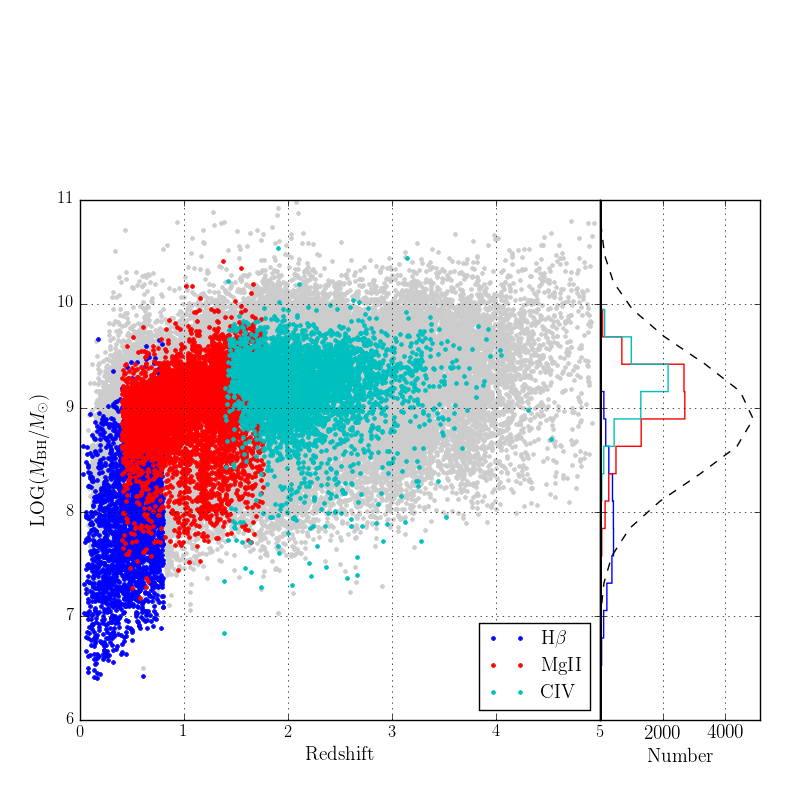} 
\caption{The left panel is the black hole mass distribution along the redshift. The grey dots are quasars from SDSS DR7 \citep{Shen2011}. The blue dots are LAMOST quasars with H${\sc \beta}$ deduced black hole mass. The red dots are LAMOST quasars with Mg{\sc ii} deduced black hole mass. The cyan dots are LAMOST quasars with C{\sc iv} deduced black hole mass. It is clear that LAMOST quasars occupy the same black hole mass space as SDSS DR7 quasars. The black mass estimate in this work is reasonable.  The right panel shows the histograms of BH masses deduced by H${\sc \beta}$ (in blue), Mg{\sc ii} (in red) and C{\sc iv} (in cyan). The black dashed line is the distribution of the BH masses for the entire DR2 and 3 quasars. \label{Fig:mbh}}
\end{figure}

In principle, the virial black hole mass should be estimated using single-epoch spectra, which is not possible for LAMOST quasars. Because the time difference between SDSS photometry and the LAMOST observation, the variations of quasar luminosity might be 0.1-0.2 mag \citep{VandenBerk2004, Zuo2012, Ai2013}. We compared our estimated BH masses to \cite{Shen2011} and show the results on the right side of Figure \ref{Fig:comp-bol-BH}.  In average, LAMOST DR2\&3 H${\sc\beta}$ estimated BH mass is $0.23\pm0.37$ dex smaller than \cite{Shen2011}, the Mg{\sc ii} estimated BH mass is $0.12\pm0.38$ dex smaller, and the C{\sc iv} estimated BH mass is $0.020\pm0.35$ dex smaller. Comparing to the FWHM differences in Figure \ref{Fig:comp-fwhm-ew}, the difference between BH mass estimate is mostly due to the difference between the FWHM measurements. We drew the BH mass distributions of LAMOST DR2\&3 and \cite{Shen2011} along the redshift in Figure \ref{Fig:mbh}. It is clear that even with a slightly different BH mass estimate, the LAMOST quasars still occupy the same BH mass space as \cite{Shen2011}. The BH mass given in this work can be considered as a good approximation. 

\section{Catalog parameters}

We provide a compiled catalog for all quasars observed in LAMOST DR2 and DR3. Parameters included in the quasar catalog is list in Table \ref{Tab:colname}. Below we describe each parameter in details.

1. OBSID: unique object identification in LAMOST database.

2. Designation: LAMOST Jhhmmss.s+ddmmss.s (J2000)

3-4. Right ascension and Declination

5-8: Spectral observation information. MJD: modified Julian data; PLANID: spectroscopic plan identification; SPID: spectrograph identification; FIBERID: fibre number. A LAMOST spectrum is named as {\it spec-MJD-PLANID-SPID-FIBERID.fits}

9: SOURCE\_FLAG: SOURCE\_FLAG is a 7-bit binary digit. The first three bits indicate whether the quasar is already reported by SDSS or NED (i.e., \textquotedblleft 011\textquotedblright\,in SDSS DR7, DR9 or DR10, \textquotedblleft 001\textquotedblright\,in DR12, \textquotedblleft 100\textquotedblright\, reported in DR14, \textquotedblleft 010\textquotedblright\, reported in NED but not in SDSS , and \textquotedblleft 000\textquotedblright\,neither reported by SDSS nor NED). The two bits in the middle indicate how the quasar candidate is selected (i.e., \textquotedblleft 00\textquotedblright\,infrared-optical colour selected, \textquotedblleft 01\textquotedblright\,data mining selected, \textquotedblleft 10\textquotedblright\,object not included in LAMOST LEGAS quasar survey sample but identified as quasar). The last two bits indicate whether the quasar is overlapped with M31/M33 field (\textquotedblleft 00\textquotedblright\,not overlapped, \textquotedblleft 01\textquotedblright\,object already reported in \cite{Huo2015}, \textquotedblleft 10\textquotedblright\,object in M31/M33 field but not in \cite{Huo2015}). For example SOURCE\_FLAG=64. \textquotedblleft 64\textquotedblright\,written in binary system is \textquotedblleft 1000000\textquotedblright\,. It means the quasar has already been reported by DR14, it was selected via infrared-optical colour selection, and is not within M31/M33 field. 

10: $M_i$. Absolute $i$ band magnitude with $K-$correction to $z=2$ following \cite{Richards2006}

11-12: CLASS\_VI and CLASS\_FLAG. Visually inspected spectral type. If the spectral type is uncertain CLASS\_FLAG = 1.

13-15:Redshift. Z\_PIPELINE: redshift given by LAMOST pipeline;  Z\_VI: redshift confirmed by visual inspection. If the value of redshift is uncertain (e.g., only one emission line available), Z\_FLAG=1. If the spectrum is likely a quasar but too noise to yield a redshift, Z\_FLAG=9. 

16: SNR\_SPEC. Median S/N per pixel in the wavelength regions of 4000-5700${\rm\AA}$ and 6200-9000${\rm\AA}$.

17-23: FWHM and the rest-frame equivalent width of H${\sc \alpha}$, [N{\sc ii}]6584, and [S{\sc ii}]$\lambda\lambda$6718, 6732

24: Rest-frame equivalent width of iron emission at 6000-6500${\rm\AA}$.

25-26: Number of good pixels and median S/N per pixel for the H${\sc \alpha}$ region in 6400-6755 ${\rm\AA}$.

27: LINE\_REDCHI2\_HA. Reduced $\chi^2$ of H${\sc \alpha}$ emission line fit.

28: LINE\_FLAG\_HA. Flag indicates the quality of H${\sc \alpha}$ fitting. As mentioned in Section \ref{em-fit}, the MPFIT package \citep{Markwardt2009} is used to perform $\chi^2$-minimization using the Levenberg-Marquardt technique. The fitting automatically assigns \textquotedblleft 0\textquotedblright\, to those with converged results and \textquotedblleft -1\textquotedblright\, without converged results. The un-converged fits are usually caused by low S/N, too few good pixels in the fitting region, or peculiar continuum and emission line properties. After fitting, we visually inspect each fitting result with LINE\_FLAG\_HA=0. For those with clearly over-subtracted continuum we changed their flag from 0 to 1 to flag an unreliable fit. If H${\sc \alpha}$ is outside the observational frame, LINE\_FLAG\_HA=-1.

29-34: FWHM and the rest-frame equivalent width of H${\sc \beta}$ and  [O{\sc iii}]$\lambda\lambda$4959, 5007. 

35: Rest-frame equivalent width of iron emission at 4435-4685 ${\rm\AA}$

36-37: Number of good pixels and median S/N per pixel for the H${\sc \beta}$ region in 6400-6755 ${\rm\AA}$.

38: LINE\_REDCHI2\_HB. Reduced $\chi^2$ of H${\sc \beta}$ emission line fit.

39: LINE\_FLAG\_HB. Quality flag for H${\sc \beta}$ fitting. It is defined using the same scheme as LINE\_FLAG\_HA.  

40-48: FWHM and the rest-frame equivalent width of Mg{\sc ii}.

49-50: Number of good pixels and median S/N per pixel for Mg{\sc ii}.

51: LINE\_REDCHI2\_MG. Reduced $\chi^2$ of Mg{\sc ii} emission line fit.

52: LINE\_FLAG\_MG. Quality flag for Mg{\sc ii} fitting. It is defined using the same scheme as LINE\_FLAG\_HA.  

53-58: FWHM and the rest-frame equivalent width for C{\sc iv}.

59-60: Number of good pixels and median S/N per pixel for C{\sc iv}.

61: LINE\_REDCHI2\_CIV. Reduced $\chi^2$ of C{\sc iv} emission line fit.

62: LINE\_FLAG\_CIV. Quality flag for C{\sc iv} fitting. It is defined using the same scheme as LINE\_FLAG\_HA. For those BAL candidates, values of LINE\_FLAG\_CIV are also changed to 1.

63: ALPHA1. Continuum spectral index $\alpha_\lambda$, where $1300<\lambda<4661\AA$

64: ALPHA2. Continuum spectral index $\alpha_\lambda$, where $\lambda>4661 \AA$

65: MODEL\_PHOT\_CHI2. Reduced $\chi^2$ of continuum fit.

66: CONT\_FLAG. Continuum fitting flag. As mentioned in Section \ref{cont-fit}, Python PYTOOLS.NMPFIT is used to perform $\chi^2$-minimization with the Levenberg-Marquardt algorithm. When the fitting converges, CONT\_FLAG=0. If the fitting cannot converge, CONT\_FLAG=1. The un-converged fitting usually happens when the SDSS 5-band magnitudes form a peculiar shape, therefore, cannot be fitted with the composite spectrum.

67-72: Monochromatic luminosity calculated from the break power-law continuum. The relative errors are calculated via Monte-Carlo simulation. 

73-81: Virial BH mass, error and flags. The BH mass error are calculated using Monte-Carlo simulation. Since BH mass is estimated using FWHM and monochromatic continuum luminosity, the flags of the BH mass are the combinations of the emission line fitting flags and the continuum fitting flags as shown in Table \ref{Tab:BH-flag}. When the LOGMBH\_FLAG = -1, there is no available BH mass estimate. If the LOGMBH\_FLAG = 1, the BH mass estimate should be used with caution. 

\begin{longtable}{|c|c|c|c|}
\caption{The determination of BH mass flags. \label{Tab:BH-flag}}\\
\hline
\multicolumn{2}{|c|}{\multirow{2}{*}{\small LOGMBH\_FLAG}} &\multicolumn{2}{|c|}{\small CONT\_FLAG}\\
\cline{3-4}
\multicolumn{1}{|c}{} & \multicolumn{1}{c|}{}& \multicolumn{1}{|c|}{1} & \multicolumn{1}{|c|}{0} \\
\cline{1-4}

\multirow{3}{*}{\begin{sideways}{\small LINE\_FLAG}\end{sideways}}& -1 & -1 & -1 \\
\cline{2-4}
& 1 & -1 & 1 \\
\cline{2-4}
& 0 & -1 & 0 \\
\hline
\end{longtable}

\section{Summary}

\begin{figure}
\includegraphics[scale=0.4]{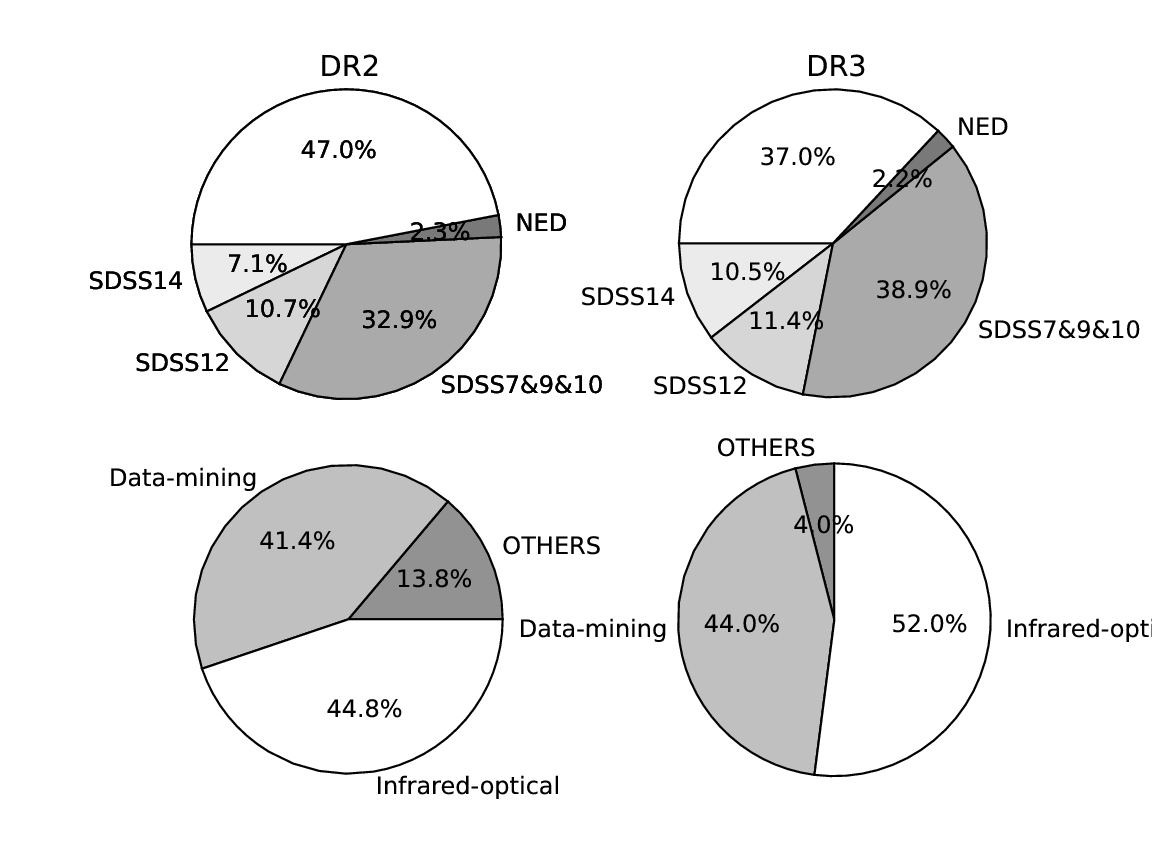} 
\caption{Sources of the quasar candidates. The top two pie plots are results from cross-matching SDSS quasars with LAMOST quasars. The grey areas are quasars reported in SDSS and NED. The white region is for quasars not reported in SDSS nor in NED. Over all, about 40\% of quasars are not reported by SDSS. It indicates that LAMOST Quasar Survey provides a great supplement to low-to-moderate redshift quasar survey. The bottom two pie plots are sources of LAMOST quasar candidates. The white region is for quasars selected by infrared-optical colour. The light-grey area is for quasars selected by data-mining. The dark-grey area is for objects that not selected as quasar candidates, but are confirmed as quasars. It shows that infrared-optical colour selection results more quasars than any other selection methods. \label{Fig:pie}}
\end{figure}

In this work, we present LAMOST LEGAS Quasar Survey Data Release Two and Three. There are 9024 confirmed quasars in DR2 and 10911 in DR3. Among them 8100 quasars are not reported in either SDSS or NED. As mentioned before, our survey candidates were finalised before SDSS DR12, therefore 2225 SDSS DR12 QSO catalogue released quasars and 1810 DR14 QSOs should also be considered as independent discovery. The bottom two pie plots in Figure \ref{Fig:pie} show the sources of the quasar candidates. In DR2, 44.8\% quasars are from infrared-optical color selection, 41.4\% are from data-mining, the remaining 13.8\% are quasars that were not submitted as quasar candidates initially. In DR3, 52.0\% quasars are from infrared-optical color selection, 44.0\% are from data-mining, the remaining 4\% are not quasar candidates but are confirmed as quasars. It indicates that infrared-optical color selection is a promising method for selecting quasars. The top two pie plots are the cross-matching results between LAMOST quasar and SDSS quasar. In both DR2 and DR3, about 50\% of LAMOST quasars are not reported by SDSS. It shows that LAMOST quasars provide a great supplement to low-to-moderate redshift quasars.

For each confirmed quasar, we applied emission line measurements of H${\sc \alpha}$, H${\sc \beta}$, Mg{\sc ii} and C{\sc iv}. We conducted a continuum fit using SDSS photometry data. We also estimated the black hole mass for each quasar. The results are compiled into one quasar catalog. Proper flags are added to indicate the liability of the fit or the estimate. Both LAMOST quasar spectra and the quasar catalog are available on-line. Follow-up studies based on LAMOST Quasar Survey, such as BAL study, are under way.

\section*{Acknowledgements}
We acknowledge the support from the National Key Basic Research Program of China (grant 2014CB845700), the Ministry of Science and Technology of China under grant 2016YFA0400703, NSFC grants 11373008 and 11533001, and the Strategic Priority Research Program "The Emergence of Cosmological Structures Grant No. XDB09000000) of National Astronomical Observatories, Chinese Academy of Sciences. The Guo Shou Jing Telescope (the Large Sky Area Multi-Object Fiber Spectroscopic Telescope,  LAMOST) is a National Major Scientific Project built by the Chinese Academy of Sciences. Funding for the project has been provided by the National Development and Reform Commission. LAMOST is operated and managed by National Astronomical Observatories, Chinese Academy of Sciences. 



\clearpage

\clearpage


\begin{longtable}{cccp{6 cm}}
\caption{The LAMOST quasar survey DR2 and 3 quasar catalog: table description \label{Tab:colname}}\\
\hline\hline
Column & Name & Format & Description \\
\hline
1& OBSID & STRING & unique object id in LAMOST database\\
2 & DESIGNATION & STRING & $hhmmss.ss+ddmmss.s$ (J2000.0)\\
3 & RA & DOUBLE & Right ascension in decimal degrees (J2000.)\\
4 & DEC & DOUBLE & Declination in decimal degrees (J2000.0)\\
5 & PLANID & STRING & spectroscopic plan identification\\
6 & SPID & STRING & spectrograph identification\\
7 & FIBERID & STRING & fibre number of the spectrum\\
8 & MJD& STRING & MJD of the spectroscopic observation\\
9 & SOURCE\_FLAG & LONG & sources of quasar candidates \\
10 & $M_i$ & DOUBLE & absolute $i$-band magnitude with $K$-correction to $z=2$ follows \cite{Richards2006}\\
\hline
11 & CLASS\_VI & STRING & classification given by visual inspection\\
12 & CLASS\_FLAG & SHORT & classification flag\\
13 & Z\_PIPELINE & DOUBLE & redshift given by LAMOST pipeline\\
14 & Z\_VI & DOUBLE & redshift given by visual inspection\\
15 & Z\_FLAG & SHORT & redshift quality flag\\
16 & SNR\_SPEC & DOUBLE & Median S/N per pixel of the spectrum\\
\hline
17 & FWHM\_BROAD\_HA & DOUBLE & FWHM of broad H{\sc $\alpha$} in km s$^{-1}$\\
18 & EW\_BROAD\_HA & DOUBLE & rest-frame equivalent width of broad H{\sc $\alpha$} in ${\rm\AA}$\\
19 & FWHM\_NARROW\_HA & DOUBLE & FWHM of narrow H{\sc $\alpha$} in km s$^{-1}$\\
20 & EW\_NARROW\_HA & DOUBLE & rest-frame equivalent width of narrow H{\sc $\alpha$} in ${\rm\AA}$\\
21 & EW\_NII\_6585 & DOUBLE & rest-frame equivalent width of [N\,{\sc ii}]6584 in ${\rm\AA}$\\
22 & EW\_SII\_6718 & DOUBLE & rest-frame equivalent width of [S\,{\sc ii}]6718 in ${\rm\AA}$\\
23 & EW\_SII\_6732 & DOUBLE & rest -frame equivalent width of [S\,{\sc ii}]6732 in ${\rm\AA}$\\
24 & EW\_FE\_HA & DOUBLE & rest -frame equivalent width of Fe within 6000-6500 ${\rm\AA}$ in ${\rm\AA}$ \\
25 & LINE\_NPIX\_HA & LONG & number of good pixels at the rest-frame 6400-6765 ${\rm\AA}$\\
26 & LINE\_MED\_SN\_HA & DOUBLE & median S/N per pixel at the rest-frame 6400-6765 ${\rm\AA}$\\
27 & LINE\_REDCHI2\_HA & DOUBLE & reduced $\chi^2$ of H{\sc $\alpha$} emission line fit\\
28 & LINE\_FLAG\_HA & SHORT & quality flag of H{\sc $\alpha$} fitting\\
\hline
29 & FWHM\_BROAD\_HB & DOUBLE & FWHM of broad H{\sc $\beta$} in km s$^{-1}$\\
30 & EW\_BROAD\_HB & DOUBLE & rest-frame equivalent width of broad H{\sc $\beta$} in ${\rm\AA}$\\
31 & FWHM\_NARROW\_HB & DOUBLE & FWHM of narrow H{\sc $\beta$} in km s$^{-1}$\\
32 & EW\_NARROW\_HB & DOUBLE & rest-frame equivalent width of narrow H{\sc $\beta$} in ${\rm\AA}$\\
33 & EW\_OIII\_4959 & DOUBLE & rest-frame equivalent width of [O\,{\sc iii}]4959 in ${\rm\AA}$\\
34 & EW\_OIII\_5007 & DOUBLE & rest-frame equivalent width of [O\,{\sc iii}]5007 in ${\rm\AA}$\\
35 & EW\_FE\_HB & DOUBLE & rest-frame equivalent width of Fe within 4435-4685 ${\rm\AA}$ in ${\rm\AA}$\\
36 & LINE\_NPIX\_HB & LONG & number of good pixels at the rest-frame 4750-4950 ${\rm\AA}$\\
37 & LINE\_MED\_SN\_HB & DOUBLE & median S/N per pixel at the rest-frame 4750-4950 ${\rm\AA}$\\
38 & LINE\_REDCHI2\_HB & DOUBLE & reduced $\chi^2$ of H{\sc $\beta$} emission line fit \\
39 & LINE\_FLAG\_HB & SHORT & quality flag of H{\sc $\beta$} fitting\\
\hline
40 & FWHM\_BROAD\_MGII\_2796 & DOUBLE & FWHM of the broad Mg{\sc ii} 2796 in km s$^{-1}$\\
41 & EW\_BROAD\_MGII\_2796 & DOUBLE & rest-frame equivalent width of the broad Mg{\sc ii} 2796 in ${\rm\AA}$\\
42 & FWHM\_NARROW\_MGII\_2796 & DOUBLE & FWHM of the narrow Mg{\sc ii} 2796 in km s$^{-1}$\\
43 & EW\_NARROW\_MGII\_2796 & DOUBLE & rest-frame equivalent width of the narrow Mg{\sc ii} 2796 in ${\rm\AA}$\\
44 & FWHM\_BROAD\_MGII & DOUBLE & FWHM of the whole broad Mg{\sc ii} in km s$^{-1}$\\
45 & EW\_BROAD\_MGII & DOUBLE & rest-frame equivalent width of the whole broad Mg{\sc ii} in ${\rm\AA}$\\
46 & FWHM\_MGII & DOUBLE & FWHM of the whole Mg{\sc ii} emission line in km s$^{-1}$ \\
47 & EW\_MGII & DOUBLE & rest-frame equivalent width of the whole Mg{\sc ii} in ${\rm\AA}$\\
48 & EW\_FE\_MGII & DOUBLE & rest-frame equivalent width of Fe within 2200-3090 ${\rm\AA}$\\
49 & LINE\_NPIX\_MGII & LONG & number of good pixels at the rest-frame 2700-2900 ${\rm\AA}$\\
50 & LINE\_MED\_SN\_MGII & DOUBLE & median S/N per pixel at the rest-frame 2700-2900 ${\rm\AA}$\\
51 & LINE\_REDCHI2\_MGII & DOUBLE & reduced $\chi^2$ of Mg{\sc ii} emission line fit\\
52 & LINE\_FLAG\_MGII & SHORT & quality flag of Mg{\sc ii} fitting\\
\hline
53 & FWHM\_BROAD\_CIV & DOUBLE & FWHM of the broad C{\sc iv} in km s$^{-1}$\\
54 & EW\_BROAD\_CIV & DOUBLE & rest frame equivalent width of the broad C{\sc iv} in ${\rm\AA}$\\
55 & FWHM\_NARROW\_CIV & DOUBLE & FWHM of the narrow C{\sc iv} in km s$^{-1}$\\
56 & EW\_NARROW\_CIV & DOUBLE & rest frame equivalent width of the narrow C{\sc iv} in ${\rm\AA}$\\
57 & FWHM\_CIV & DOUBLE & FWHM of the whole C{\sc iv} in km s$^{-1}$\\
58 & EW\_CIV & DOUBLE &  rest-frame equivalent width of the whole C{\sc iv} in ${\rm\AA}$\\
59 & LINE\_NPIX\_CIV & LONG & number of good pixels at the rest-frame 1500-1600 ${\rm\AA}$\\
60 & LINE\_MED\_SN\_CIV & DOUBLE & median S/N per pixel at the rest-frame 1500-1600 ${\rm\AA}$\\
61 & LINE\_REDCHI2\_CIV & DOUBLE & reduced $\chi^2$ of C{\sc iv} emission line fit\\
62 & LINE\_FLAG\_CIV & SHORT & quality flag of C{\sc iv} fitting\\
\hline
63 & ALPHA1 & DOUBLE & wavelength power-law index from 1300 to 4661 ${\rm\AA}$\\
64 & ALPHA2 & DOUBLE & wavelength power-law index from 4661 ${\rm\AA}$ towards the red end of the spectrum \\
65 & MODEL\_PHOT\_CHI2 & DOUBLE & reduced chi-square of continuum fitting \\
66 & CONT\_FLAG & SHORT & quality flag of continuum fitting \\
67 & LOGL1350 & DOULBE & monochromatic luminosity at 1350 ${\rm\AA}$ in units of erg/s\\
68 & LOGL1350\_ERR & DOUBLE & error of LOGL1350 in units of erg/s \\
69 & LOGL3000 & DOUBLE & monochromatic luminosity at 3000 ${\rm\AA}$ in units of erg/s\\
70 & LOGL3000\_ERR & DOUBLE & error of LOGL3000  in units of erg/s \\
71 & LOGL5100 & DOUBLE & monochromatic luminosity at 5100 ${\rm\AA}$ in units of erg/s\\
72 & LOGL5100\_ERR & DOUBLE & error of LOGL5100 in units of erg/s \\
\hline
73 & LOGMBH\_HB & DOUBLE & virial BH mass based on H{\sc $\beta$} in units of $M_\odot$ \\
74 & LOGMBH\_ERR\_HB & DOUBLE & error of LOGMBH\_HB in units of $M_\odot$ \\
75 & LOGMBH\_HB\_FLAG & SHORT & reliability of LOGMBH\_HB\\
76 & LOGMBH\_MGII & DOUBLE & virial BH mass based on Mg{\sc ii} in units of $M_\odot$\\
77 & LOGMBH\_ERR\_MGII & DOUBLE & error of LOGMBH\_MGII in units of $M_\odot$\\
78 & LOGMBH\_MGII\_FLAG & SHORT & reliability of LOGMBH\_MGII\\
79 & LOGMBH\_CIV & DOUBLE & virial BH mass based on C{\sc iv} in units of $M_\odot$\\
80 & LOGMBH\_ERR\_CIV & DOUBLE & error of LOGMBH\_CIV in units of $M_\odot$\\
81 & LOGMBH\_CIV\_FLAG & SHORT & reliability of LOGMBH\_CIV \\
\hline\hline
\end{longtable}

\clearpage


\end{document}